\documentclass[epjST]{svjour}
\usepackage{subfigure}
\usepackage{graphicx}
\usepackage{amsmath}
\usepackage{amssymb}
\usepackage[sort&compress,numbers]{natbib}
\usepackage{color}

\begin{document}

\date{\today}

\title{Reducing the number of time delays in coupled dynamical systems}

\author{Alexandre Wagemakers\inst{1}\fnmsep\thanks{\email{alexandre.wagemakers@urjc.es}} \and Javier Used \inst{1} \and Miguel A. F. Sanju\'{a}n\inst{1}\inst{2}\inst{3}}

\institute{Nonlinear Dynamics, Chaos and Complex Systems Group, Departamento de  F\'isica, Universidad Rey Juan Carlos, M\'ostoles, Madrid, Tulip\'an s/n, 28933, Spain
\and
Department of Applied Informatics, Kaunas University of Technology, Studentu 50-415, Kaunas LT-51368, Lithuania
\and
Institute for Physical Science and Technology, University of Maryland, College Park, Maryland 20742, USA}


\abstract{When several dynamical systems interact, the transmission of the information between them necessarily implies a time delay. When the time delay is not negligible, the study of the dynamics of these interactions deserve a special treatment. We will show here that under certain assumptions, it is possible to reduce the number of time delays without altering the global dynamics. We will focus here on graphs of interactions with identical time delays and bidirectional connections. With these premises, it is possible to find a configuration where a number $n_z$ of time delays have been removed with $n_v-1 \leq n_z \leq n_v^2/4$, where $n_v$ is the number of dynamical systems on a connected graph.}

\maketitle
\section{Introduction}

Time delays appear in a very natural way in any communication from one entity to another. In the context of dynamical systems, it is intrinsic to the transmission of the state through a communication channel. When the time delay is very small compared to the time scale of the processes involved in the dynamical systems, the induced effects of these time delays are barely noticeable, but still present. { While delayed dynamical systems have infinite dimension, the effective motion of the trajectories evolves on a finite  dimensional manifold. When the time delay increases, the complexity of the system grows continuously with the increase of the dimension of the effective manifold \cite{yanchuk2017spatio}.}

It is easy to find examples on graphs with time delays in very different fields. In physics, such time delays have been studied for coupled semiconductor lasers \cite{soriano2013complex}. In engineering, we can cite the problem of consensus among agents on a graph with a time delay \cite{Olfati_2004}. These time delays affect the dynamics of the whole graph and makes it harder the analysis and the simulation of the system. Another example where time delays can have relevant effects on the dynamics of a graph is the communication between cells of the nervous system \cite{liang09}. The problem of the synchronization of the dynamical systems on a graph with time-delayed coupling is partially understood \cite{li_2004,yeung1999time,ott_delay}, though the analysis is difficult to achieve in most cases.

In an attempt to reduce the dimensionality of the system, the method called componentwise time-shift transformation \cite{lucken_reduction_2013} allows to transform the time delays on the graph. The method relies on an invariant of the graph: the time delays can be altered as long as the sum of the time delays on the constitutive loop of the graph remain constant. This method was used successfully to demonstrate that we can reduce the number of time delays on any graph by setting $n_z=n_v-1$ time delays to zero, where $n_v$ is the number of vertices of the graph \cite{lucken_reduction_2013,lucken_classification_2015}.

A new formulation of this method \cite{wagemakers_2017} has been developed to improve the number of zero time delays. With the application of this new method, we find a number of zero time delays $n_z\geq n_v-1$, and the total sum of the time delays is also considerably reduced.

While this last technique is useful in general, we will show here that in some special cases of interest, the number of zero time delays can be increased by a large extent. The main objective of this work is to show that when the graphs possess identical time delays and bidirectional links between pairs, the maximum number of zeros is bounded between $n_v-1 \leq n_z \leq n_v^2/4$. The lower bound corresponds to the minimum achievable on any graph, while the upper bound is the corresponding to the complete bipartite graph with the same number of vertices in each partition.

To achieve this goal, first we explain the basic techniques of the componentwise time-shift transformation that allows to move around time delays without altering the dynamics. Then, we consider the particular situation of graphs with identical time delays. We also show how finding the maximum number of zero time delays can be reduced to a combinatorial search on a graph. Finally, we use numerical simulations to test the analytical results.

\section{Componentwise time-shift transformation}

We consider a graph $G$ with a collection of $n_e$ oriented edges $e_{i}$ and $n_v$ vertices $v_i$. At each vertex we have a very general dynamical system. The equation set takes the form of a system of coupled delay differential equations
\begin{equation} \label{sys_din}
  \frac{dx_i}{dt} = f_i(x_i(t), x_j(t-\tau_{k})_{k \in S_i}),
\end{equation}
with $i=1,...,n_v$ and $S_i$ is the set of indices $k$ such that the edges $e_k$ connect the vertex $j$ to the vertex $i$. We assume a discrete time delay $\tau_{k}$ on the edge $e_k$.

The previous system in Eq.~(\ref{sys_din}) can be transformed with a redefinition of the time delays $\tau_{k}$ without changing the dynamical properties of the system using the componentwise time-shift transformation \cite{lucken_reduction_2013,lucken_classification_2015}. We set
\begin{equation} \label{sys_din_shift}
\frac{dy_i}{dt} = f_i(y_i(t), y_j(t-\tilde \tau_{k})_{k \in S_i}),
\end{equation}
with the following change of variables
\begin{align} \label{delay_shift}
y_i(t) = x_i(t-\eta_i) \\
\tilde \tau_{k} = \tau_{k} + \eta_{s(k)} - \eta_{t(k)},
\end{align}
where $\eta_i$ are constants and $s(k)$ is the source vertex of the edge $k$, and $t(k)$ the target vertex of the same edge. The authors in~\cite{lucken_reduction_2013} noticed that the {\it algebraic sum} of the time delays around any cycle of the graph is constant for every choice of the time-shifts $\eta_i$. The term {\it algebraic sum} means here that, given an oriented cycle in the graph, the time delay associated to the edges on the cycle with the same orientation should be summed up and the time delays on edges with opposite direction subtracted.

Now the problem is to find the time-shifts $\eta_i$ associated to each vertex for a desired configuration of time delays $\tilde \tau_{k}$. It is possible to find a vector {$\boldsymbol\eta$} based on the topology of the graph and the time delay requirements on each edge \cite{wagemakers_2017}. Notice that the initial history of the delay differential equation deserves a special treatment if we want to match the trajectory in the phase space for both sets of equations~\cite{lucken_classification_2015}.

\section{Time-delay reduction on bidirectional graphs}

In the following we assume several general hypotheses. First, we consider directed graphs with bidirectional edges, which means that each connected vertex has a pair of edges in both directions. We make the distinction between this case and undirected graphs since the time delays might be different depending on the direction of the edge. We will also restrict our attention to graphs with identical time delays on each edge. At last, we also assume that the graph is weakly connected. With these general considerations in mind, we show that it is possible to change the distribution of the time delays on the graph according to simple rules.

\begin{figure}
  \begin{center}
  \subfigure[]{\includegraphics{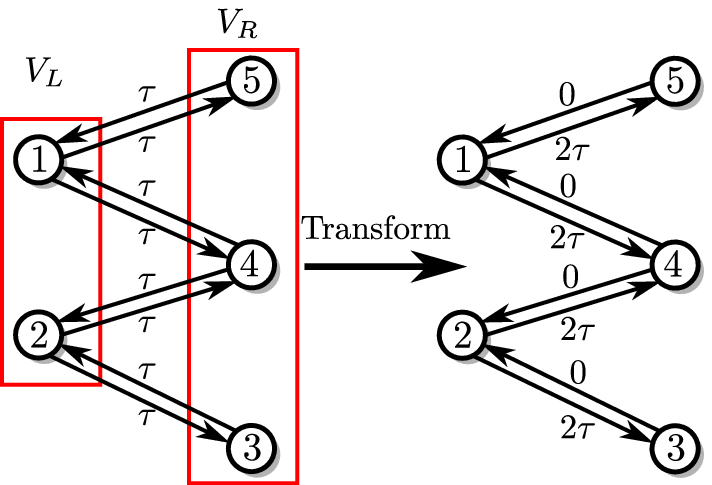}}\\
  \subfigure[]{\includegraphics{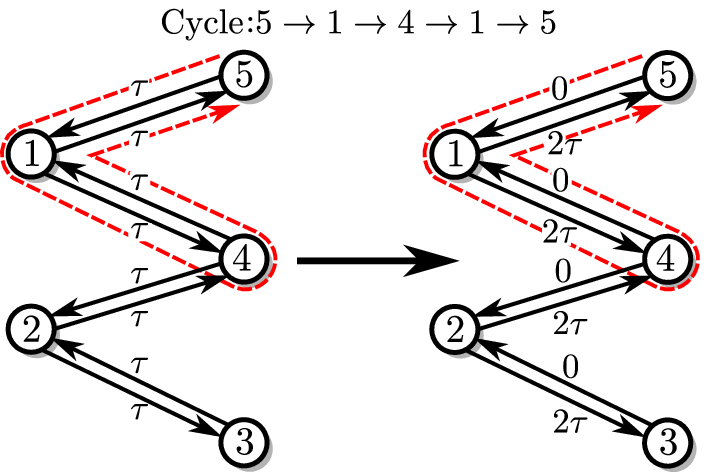}}
  \caption{This figure illustrates the process of assigning different time delays on a bipartite graph. In panel(a) the transformation sets a zero time delay on the edges from the set $V_R$ to the set $V_L$, and a time delay $2\tau$ in the direction $V_L$ to $V_R$. In panel(b) there is an example of a cycle on the graph. The sum of the time delays along the edges is the same on the original graph (on the left) and the transformed graph (on the right). The sum of time delays on the cycle $5\to 1 \to 4 \to 1 \to 5$ is $4\tau$ in both cases.}
  \label{fig1}
\end{center}
\end{figure}

To understand the method, we first limit the study to bipartite graphs. A bipartite graph has two sets of vertices such that there are only edges between these two sets; in other words, there is no edge between any two vertices of the same set. If we apply the idea of conservation of the time delay around a loop of the graph, we have noticed that a simple transformation removes half of the time delays in this graph. First, we label the two sets of nodes $V_L$ and $V_R$ depending on the set of the bipartite partition. For simplicity, we label the vertices $R$ if they belong to $V_R$, and $L$ when they belong to $V_L$. If the time delay in the graph is $\tau$, we set arbitrarily a time delay $\tau_{LR}=2\tau$ for the edges going from a vertex $L$ to $R$, and a time delay $\tau_{RL}=0$ for edges in the opposite direction. This process is illustrated in Fig. \ref{fig1}. Since any cycle on the graph has to alternate between both sets, the number of edges from $R$ to $L$, and $L$ to $R$ has to be the same. Any cycle of length $n$ has a total time delay $n\cdot\tau$, which would be conserved with the asymmetrical distribution of time delays. We claim that this is the optimal time delay distribution in the sense of the number of zeros on the edges, that is, $n_z=n_e/2$.

{This analysis leads to an interesting consequence on other graphs that do not have the bipartite property. We can reduce the number of time delays on a bipartite sub-graph without changing the sum of the time delay on the loops of the entire network. To illustrate this effect, we suppose a bipartite network with identical time delays and only one edge between two vertices of the same set, $V_R$ or $V_L$. For the purpose of the discussion, we will call this particular edge $e$. Any cycle has to run through an even number of edges between $V_L$ and $V_R$ since it has to go back and forth between the two sets before returning to its initial vertex. As a consequence, if a cycle passes once through $e$, the total sum of the time delays along the cycle will be $(n+1)\cdot\tau$ with $n$ the even number of trips between $V_L$ and $V_R$. If we reduce the bipartite graph according to the method mentioned earlier, the sum of the time delays along the cycle will remain unchanged. This invariance can be explained by noticing that the sum of the time delays along the edges between the sets $V_L$ and $V_R$ is conserved. It leaves unchanged the sum of the time delays around any cycle of the graph including the cycles passing through $e$. This reasoning can be extended to any number of edges between the vertices of the same set $V_L$ or $V_R$.}

\begin{figure}
  \begin{center}
  \subfigure[]{\includegraphics{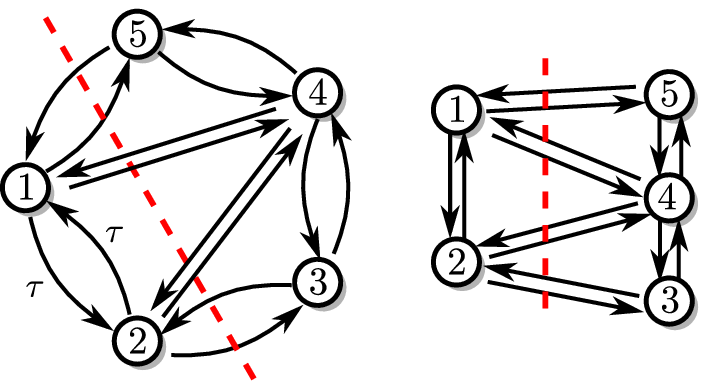}}\\
    \subfigure[]{\includegraphics{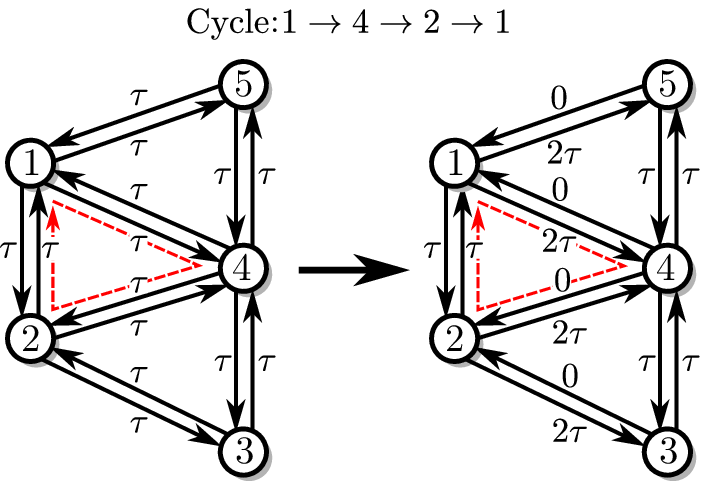}}
  \caption{Panel(a). Example of the optimal cut in a simple graph. The cut cleaves the vertices into two subsets. The number of edges between the two subsets is optimal. Panel(b). The original graph has been transformed into its equivalent asymmetric time delay distribution on the edges. The sum of the time delay over the cycle  $1 \to 4 \to 2 \to 1 $ outlined in red, is the same in the original and the transformed graph.}
  \label{fig2}
\end{center}
\end{figure}

We can now express the main result of this section. Given a graph with the general properties listed above, we can find a distribution of time delays with a number of zeros
\begin{equation}\label{theoretic_bound}
n_v-1 \leq n_z \leq n_v^2/4.
\end{equation}
We conjecture that the maximum number of zeros $n_z$ in the graph is given by the bipartite subgraph with the maximum number of edges. The upper bound corresponds to the complete bipartite subgraph with $n_v^2$, while the lower bound can be satisfied on any graph \cite{lucken_reduction_2013}.

The problem of finding the bipartite subgraph with the largest number of edges is a well-known problem in combinatorial optimization named the MAXCUT problem. The problem consists in finding a partition of the vertices that will have a maximum number of edges between the two sets. This problem is known to be computationally difficult (NP hard) but suboptimal solutions can be found in polynomial time \cite{Goemans_1995}. Figure \ref{fig2}(a) shows an example of an optimal cut in a simple graph, where the cut intersects 8 edges. In Fig.~\ref{fig2}(b), we show that the cycle passing through the vertices $1 \to 4 \to 2 \to 1 $ has the same cumulative sum of time delays around the cycle in both the original and the transformed graph.

In the next section we will show some results of the application of the optimization of graphs to get the largest bipartite graph. Once the bipartite graph has been found, we can change the distribution of time delays on the subset of edges following the method we have described earlier.

\section{Examples of time-delay reduction on graphs}

We will demonstrate the effectiveness of the method on graphs with well-known characteristics. For our purpose, the only information needed on the graph is its adjacency matrix $A$. It has dimension $n_v\times n_v$ and if an edge connects the vertex $i$ to the vertex $j$, the entry $a_{ij}$ of the adjacency matrix $A$ is $1$, and $0$ in the other case. The MAXCUT algorithm can be stated as follows
\begin{equation} \label{maxcut_pb}
\begin{array}{ll}
\textrm{Maximize: }&  \displaystyle\sum_{i,j} a_{ij} -  \frac{1}{2} c^T A c\\
&\\
\textrm{with }c_k ~ \in \{-1,1\},& \\
\end{array}
\end{equation}
where $c$ is a column vector of dimension $n_v$. This is a integer program with a quadratic objective that can be solved with an standard optimization software. The entry $c_i$ of the vector $c$ classifies the vertex $i$ into the set $V_L$ or $V_R$. Once the algorithm has found a solution, the number of edges that crosses the cut is the value
\begin{equation}
  n_c = \displaystyle\sum_{i,j} a_{ij}- \frac{1}{2} c^T A c.
\end{equation}
The number of zero time delays allowed on the graph is $n_z=n_c/2$ as described in the previous section.

\begin{figure}
  \begin{center}
  \subfigure[]{\includegraphics[height=5cm]{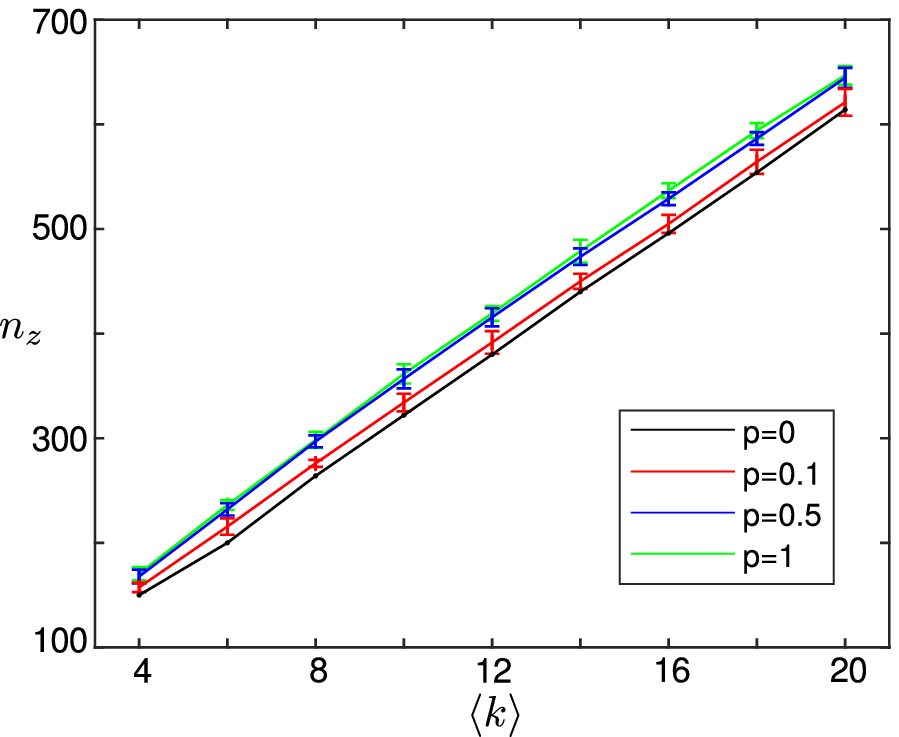}}
  \subfigure[]{\includegraphics[height=5cm]{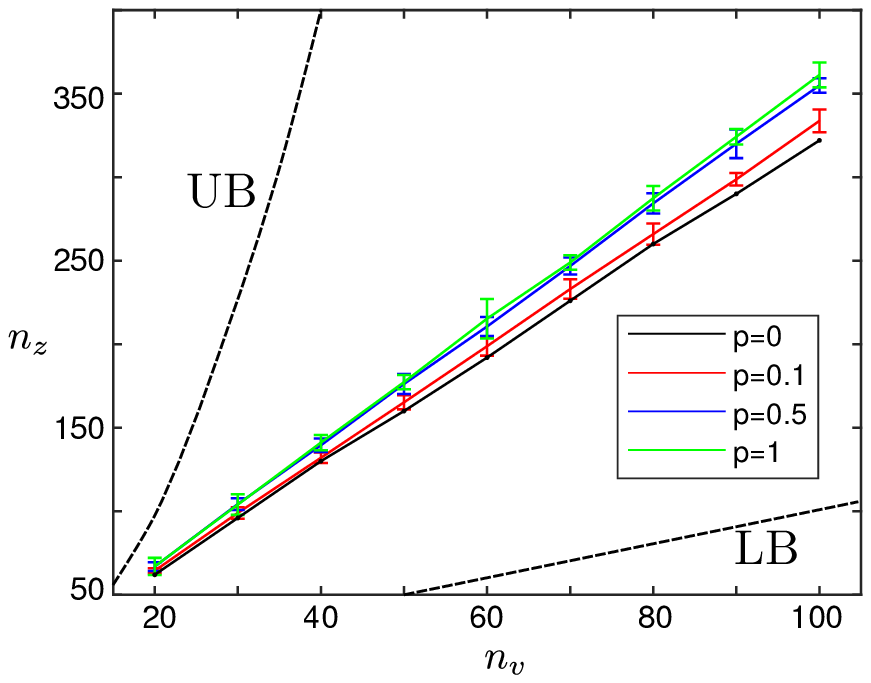}}
  \caption{Panel(a), we represent the evolution of the results of the maximum number of zero time delay $n_z$ for the Watts-Strogatz small-world model with $n_v=100$. The four curves correspond to $p=0$, $p=0.1$, $p=0.5$, and $p=1$. The MAXCUT optimization behaves linearly as a function of the vertex mean degree $\langle k\rangle$. { The results are between the theoretical bounds of Eq.~(\ref{theoretic_bound}), that is $99 \leq n_z \leq 2500$.} Panel(b) represents $n_z$ as a function of the number of vertices $n_v$ for $\langle k\rangle=10$. The evolution is clearly linear for the four cases $p=0$, $p=0.1$, $p=0.5$, and $p=1$. The lower (LB) and upper (UB) bound of Eq. (\ref{theoretic_bound}) appear in dashed line on the figure.}
  \label{fig3}
\end{center}
\end{figure}

To test the algorithm we will focus on two important models: the Watts-Strogatz small-world and the scale-free models \cite{newman2010networks}. In the small-world model, we can vary a parameter $p$ such that the graph evolves continuously from a regular graph to a random Erd\"os-Renyi graph. For $p=0$ the graph is regular so that each vertex is connected to $k$ neighbors. As the parameter $p$ increases from $0$ to $1$, some of the edges are redirected to create shortcuts in the initial symmetric configuration. When $p$ gets eventually to $1$ we have an Erd\"os-Renyi random graph where the probability to have an edge between two vertices is $p_e=n_v/n_e=1/k$.
\begin{figure}
  \begin{center}
  \includegraphics[height=4cm]{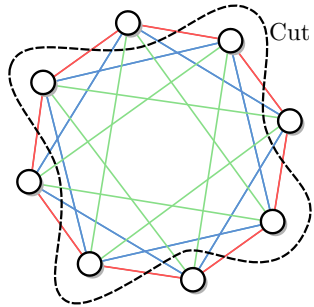}
  \caption{This figure illustrates a particular cut (in dashed line) of a regular graph where each vertex is connected to $k=6$ nearest neighbors. The cut includes alternative vertices into the same set. The number of edges crossing the cut is $n_v\cdot \left[ \frac{k}{2} \right]$, in this case $8\cdot \left[ \frac{6}{2} \right]=24$}
  \label{fig_regular_cut}
\end{center}
\end{figure}
Figure~\ref{fig3}(a) depicts the number of zero time delays $n_z$ of several generations of the graph as a function of the vertex mean degree $\langle k\rangle$. We have chosen four values of the probability $p$ of the graph: $p=0$ (regular graph), $p=0.1$, $p=0.5$, and $p=1$ (random graph). For $\langle k\rangle \geq 5$ the tendency is linear in all cases. In Fig.~\ref{fig3}(b) we have the result of the optimization as a function of the number of vertices $n_v$ for $\langle k\rangle=10$. It is remarkable that we have also a linear trend in all cases. From the analysis of the figures, we observe that the result of the optimization do not depend clearly on $p$. While we do not have an explanation for this linear behavior, we can give simple arguments based on the two extreme cases $p=0$ and $p=1$.

{When the graph is regular ($p_0$), we have found a special partition of the graph that brings an analytic formula for the number $n_c$. It simply consists in picking alternate neighbor vertices to form the partition as shown in Fig.~\ref{fig_regular_cut}. If we count the number of edges from one partition to the other, and using symmetry arguments, we get}
\begin{equation}\label{nc_SW}
n_c =  n_v\cdot \left[ \frac{k}{2} \right],
\end{equation}
where $[x]$ represents the closest integer higher or equal to $x$; notice that this formula works for even and odd number of vertices. It seems however that this cut is not optimal. In many cases the MAXCUT algorithm brings a better solution. Given that $n_z=n_c/2$, it provides us a simple lower bound for the number of zero time delays in the small-world model with $p=0$. For the case $p=1$, we can partially explain the results by considering a random cut that separates the vertices in two sets with the same number of elements. { To achieve this, we just randomly pick half of the vertices to form either the set $V_L$ or $V_R$ and we begin to construct a random Erd\"os-Renyi graph. Each vertex from the set $V_L$ has approximately $n_v/2$ vertices to choose from the set $V_R$, so that there are $(n_v/2)^2$ possible edges. Since the probability to form an edge between the two vertices is $p\simeq \langle k\rangle / n_v$, and the formation is independent for each edge, the average number of edges that will cross the cut is}
\begin{equation}
  \langle n_c \rangle = \frac{\langle k\rangle}{n_v}\cdot\frac{ n_v^2}{4}  =\frac{ n_v \cdot \langle k\rangle }{4}.
\end{equation}
This is in fact linear with $n_v$ and the mean vertex degree $\langle k\rangle$. The MAXCUT algorithm finds a better solution than this naive random cut, but the linearity of the graph $n_z$ against $n_v$ remains. For other values of $p$, it is not obvious how to obtain $n_c$ and we need numerical simulations to obtain an estimation. While the parameter $p$ has a tremendous effect on the graph theoretic properties, it seems that it has a very limited effect on the number $n_z$. This is a surprising finding that indicates that topological factors such as the graph diameter and the shortest path length have little relevance here.

In the last example, we show some results on the scale-free networks, another paradigmatic topology of graphs. The graphs are constructed following the preferred attachment algorithm and we have optimized the graph following the same method. We compare three cases: scale-free model, small-world with $p=0.5$, and Erd\"os-Renyi (small-world with $p=1$).

\begin{figure}
  \begin{center}
  \subfigure[]{\includegraphics[height=5cm]{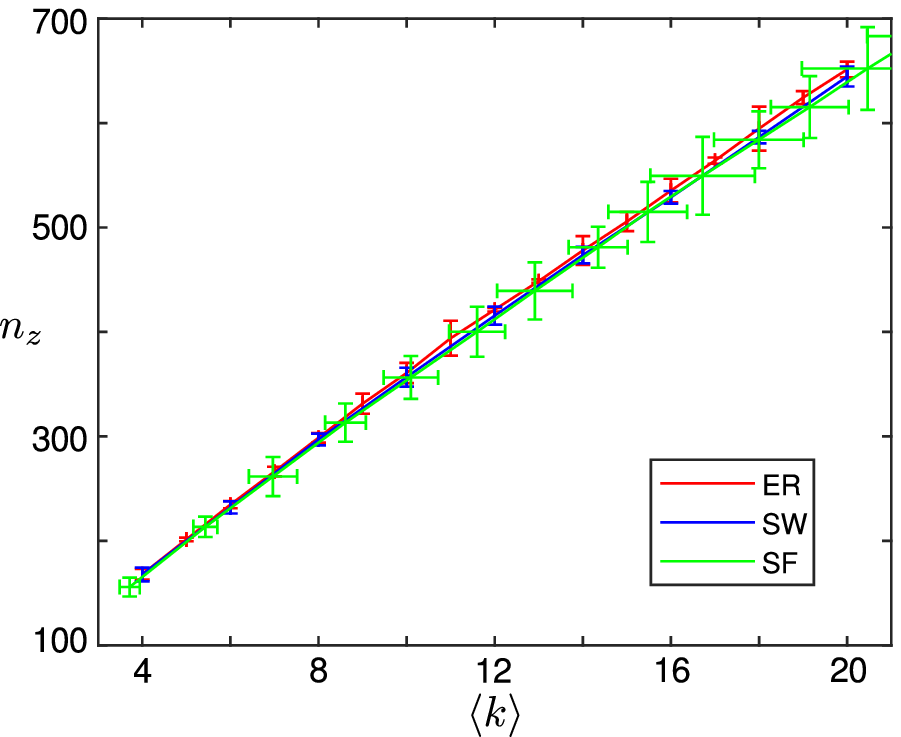}}
  \subfigure[]{\includegraphics[height=5cm]{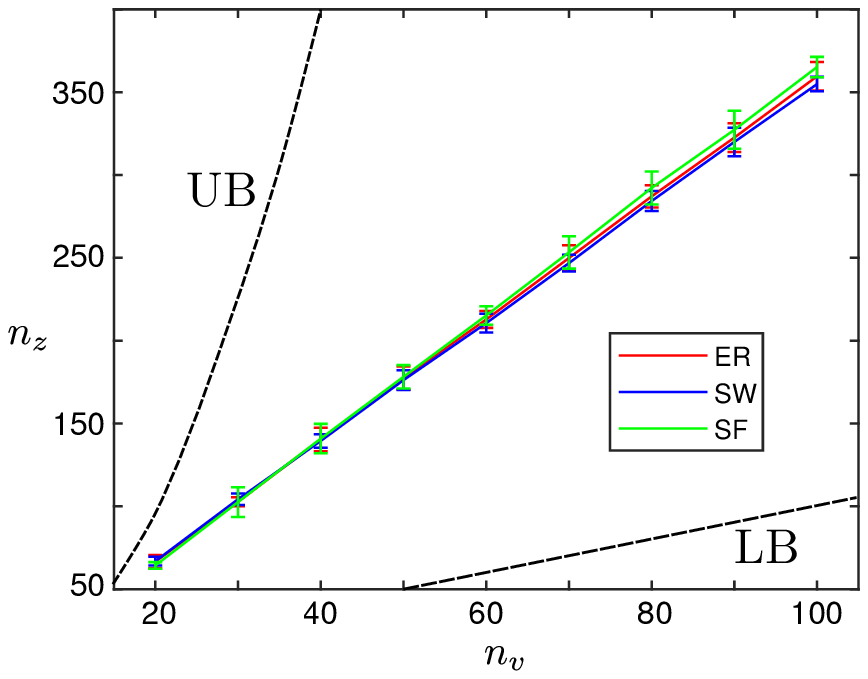}}
  \caption{In panel (a), we represent the evolution of the results of the maximum number of zero time delays $n_z$ for three different models: scale-free (SF), small-world (SW) with $p=0.5$ and Erd\"os-Renyi (ER) with $n_v=100$. The MAXCUT optimization behaves linearly as a function of the vertex mean degree $\langle k\rangle$. Panel (b) represents $n_z$ as a function of the number of vertices $n_v$ for the three models for $\langle k\rangle=10$. Notice that the three models have almost the same maximum number of zeros independently of the topology. The lower (LB) and upper (UB) bound of Eq. (\ref{theoretic_bound}) appear in dashed line on the figure.}
  \label{fig4}
\end{center}
\end{figure}

The results shown in Fig.~\ref{fig4} for $n_z$ are almost identical for the three cases. The MAXCUT algorithm does not depend on the chosen type of topology, but a more comprehensive study is required to understand the reason of this similarity.

{ The numerical simulations have been performed with the packages Lightgraphs and JuMP of the Julia programming language along with the IBM optimization software CPLEX for the solution of the integer program.}

\section{Discussion and Conclusions}

When the graph has an identical distribution of time delays, it is possible to find a new distribution of the time delays on the graph so that the global dynamics is not affected. The MAXCUT algorithm finds a bipartite subgraph with the maximum number of edges between the two sets. This partition is the basis for the new distribution of the time delays. The method basically consists in assigning on the edges connecting the two sets, zero time delays in one direction, and twice the initial time delay in the opposite direction.

We have established that according to this method, the maximum number of zero time delays in the graph that can be achieved is $n_v^2 /4 $, while the minimum is $n_v-1$. On the basis of the simulations of the last section, we can conjecture a lower bound much tighter for the special type of graphs that we are studying
\begin{equation}
n_z \geq \frac{n_v\cdot k}{4}.
\end{equation}
It is possible to reach the upper bound $n_z=n_v^2 /4 $ when there is a complete bipartite subgraph embedded in the original graph.

The numerical simulations also indicate that the topology is not a critical factor for the result of the MAXCUT optimization. It seems that the average node degree and the number of vertices are the two key parameters that affect $n_z$. It would be interesting on its own to study more in detail the result of the MAXCUT algorithm as a function of different topological factors of the graph.

Delay differential equations are in general very difficult to analyze and simulate. We have shown here that it is possible to significantly reduce the dimensionality of a set of coupled delay differential equations over a bidirectional graph with identical time delays. The reduction is at least of the order of the number of vertices. The process of changing the time delays over the graph is simple, but it is hard to find a suitable partition of the nodes. Fortunately, for large graphs there are suboptimal approximations to the MAXCUT algorithm bringing a solution in a polynomial time. Furthermore, this method could be extended to directed graphs with identical time delay.

This work opens new perspectives on the simulation and the analysis of large systems of delay coupled dynamical systems. We believe that this method will bring some insight on the behavior of collective dynamics with time delay.

\section*{Acknowledgements}

This work was supported by the Spanish State Research Agency (AEI) and the European Regional Development Fund (FEDER) under Project No.~FIS2016-76883-P. MAFS acknowledges the jointly sponsored financial support by the Fulbright Program and the Spanish Ministry of Education (Program No. FMECD-ST-2016).


\begin{thebibliography}{12}%
\makeatletter
\providecommand \@ifxundefined [1]{%
 \@ifx{#1\undefined}
}%
\providecommand \@ifnum [1]{%
 \ifnum #1\expandafter \@firstoftwo
 \else \expandafter \@secondoftwo
 \fi
}%
\providecommand \@ifx [1]{%
 \ifx #1\expandafter \@firstoftwo
 \else \expandafter \@secondoftwo
 \fi
}%
\providecommand \natexlab [1]{#1}%
\providecommand \enquote  [1]{``#1''}%
\providecommand \bibnamefont  [1]{#1}%
\providecommand \bibfnamefont [1]{#1}%
\providecommand \citenamefont [1]{#1}%
\providecommand \href@noop [0]{\@secondoftwo}%
\providecommand \href [0]{\begingroup \@sanitize@url \@href}%
\providecommand \@href[1]{\@@startlink{#1}\@@href}%
\providecommand \@@href[1]{\endgroup#1\@@endlink}%
\providecommand \@sanitize@url [0]{\catcode `\\12\catcode `\$12\catcode
  `\&12\catcode `\#12\catcode `\^12\catcode `\_12\catcode `\%12\relax}%
\providecommand \@@startlink[1]{}%
\providecommand \@@endlink[0]{}%
\providecommand \url  [0]{\begingroup\@sanitize@url \@url }%
\providecommand \@url [1]{\endgroup\@href {#1}{\urlprefix }}%
\providecommand \urlprefix  [0]{URL }%
\providecommand \Eprint [0]{\href }%
\providecommand \doibase [0]{http://dx.doi.org/}%
\providecommand \selectlanguage [0]{\@gobble}%
\providecommand \bibinfo  [0]{\@secondoftwo}%
\providecommand \bibfield  [0]{\@secondoftwo}%
\providecommand \translation [1]{[#1]}%
\providecommand \BibitemOpen [0]{}%
\providecommand \bibitemStop [0]{}%
\providecommand \bibitemNoStop [0]{.\EOS\space}%
\providecommand \EOS [0]{\spacefactor3000\relax}%
\providecommand \BibitemShut  [1]{\csname bibitem#1\endcsname}%
\let\auto@bib@innerbib\@empty
\bibitem [{\citenamefont {Yanchuck}\ and\ \citenamefont
  {Giacomelli}(2017)}]{yanchuk2017spatio}%
  \BibitemOpen
  \bibfield  {author} {\bibinfo {author} {\bibfnamefont {S.}~\bibnamefont
  {Yanchuck}}\ and\ \bibinfo {author} {\bibfnamefont {G.}~\bibnamefont
  {Giacomelli}},\ }\href@noop {} {\bibfield  {journal} {\bibinfo  {journal} {J.
  Phys. A: Math. Theor.}\ }\textbf {\bibinfo {volume} {50}},\ \bibinfo {pages}
  {103001} (\bibinfo {year} {2017})}\BibitemShut {NoStop}%
\bibitem [{\citenamefont {Soriano}\ \emph {et~al.}(2013)\citenamefont
  {Soriano}, \citenamefont {Garc{\'\i}a-Ojalvo}, \citenamefont {Mirasso},\ and\
  \citenamefont {Fischer}}]{soriano2013complex}%
  \BibitemOpen
  \bibfield  {author} {\bibinfo {author} {\bibfnamefont {M.~C.}\ \bibnamefont
  {Soriano}}, \bibinfo {author} {\bibfnamefont {J.}~\bibnamefont
  {Garc{\'\i}a-Ojalvo}}, \bibinfo {author} {\bibfnamefont {C.~R.}\ \bibnamefont
  {Mirasso}}, \ and\ \bibinfo {author} {\bibfnamefont {I.}~\bibnamefont
  {Fischer}},\ }\href@noop {} {\bibfield  {journal} {\bibinfo  {journal} {Rev.
  Mod. Phys.}\ }\textbf {\bibinfo {volume} {85}},\ \bibinfo {pages} {421}
  (\bibinfo {year} {2013})}\BibitemShut {NoStop}%
\bibitem [{\citenamefont {Olfati-Saber}\ and\ \citenamefont
  {Murray}(2004)}]{Olfati_2004}%
  \BibitemOpen
  \bibfield  {author} {\bibinfo {author} {\bibfnamefont {R.}~\bibnamefont
  {Olfati-Saber}}\ and\ \bibinfo {author} {\bibfnamefont {R.~M.}\ \bibnamefont
  {Murray}},\ }\href@noop {} {\bibfield  {journal} {\bibinfo  {journal} {IEEE
  Transactions on Automatic Control}\ }\textbf {\bibinfo {volume} {49}},\
  \bibinfo {pages} {1520} (\bibinfo {year} {2004})}\BibitemShut {NoStop}%
\bibitem [{\citenamefont {Liang}\ \emph {et~al.}(2009)\citenamefont {Liang},
  \citenamefont {Tang}, \citenamefont {Dhamala},\ and\ \citenamefont
  {Liu}}]{liang09}%
  \BibitemOpen
  \bibfield  {author} {\bibinfo {author} {\bibfnamefont {X.}~\bibnamefont
  {Liang}}, \bibinfo {author} {\bibfnamefont {M.}~\bibnamefont {Tang}},
  \bibinfo {author} {\bibfnamefont {M.}~\bibnamefont {Dhamala}}, \ and\
  \bibinfo {author} {\bibfnamefont {Z.}~\bibnamefont {Liu}},\ }\href@noop {}
  {\bibfield  {journal} {\bibinfo  {journal} {Phys. Rev. E}\ }\textbf {\bibinfo
  {volume} {80}},\ \bibinfo {pages} {066202} (\bibinfo {year}
  {2009})}\BibitemShut {NoStop}%
\bibitem [{\citenamefont {Li}\ and\ \citenamefont {Chen}(2004)}]{li_2004}%
  \BibitemOpen
  \bibfield  {author} {\bibinfo {author} {\bibfnamefont {C.}~\bibnamefont
  {Li}}\ and\ \bibinfo {author} {\bibfnamefont {G.}~\bibnamefont {Chen}},\
  }\href@noop {} {\bibfield  {journal} {\bibinfo  {journal} {Physica A:
  Statistical Mechanics and its Applications}\ }\textbf {\bibinfo {volume}
  {343}},\ \bibinfo {pages} {263 } (\bibinfo {year} {2004})}\BibitemShut
  {NoStop}%
\bibitem [{\citenamefont {Yeung}\ and\ \citenamefont
  {Strogatz}(1999)}]{yeung1999time}%
  \BibitemOpen
  \bibfield  {author} {\bibinfo {author} {\bibfnamefont {M.~S.}\ \bibnamefont
  {Yeung}}\ and\ \bibinfo {author} {\bibfnamefont {S.~H.}\ \bibnamefont
  {Strogatz}},\ }\href@noop {} {\bibfield  {journal} {\bibinfo  {journal}
  {Phys. Rev. Lett.}\ }\textbf {\bibinfo {volume} {82}},\ \bibinfo {pages}
  {648} (\bibinfo {year} {1999})}\BibitemShut {NoStop}%
\bibitem [{\citenamefont {Lee}\ \emph {et~al.}(2009)\citenamefont {Lee},
  \citenamefont {Ott},\ and\ \citenamefont {Antonsen}}]{ott_delay}%
  \BibitemOpen
  \bibfield  {author} {\bibinfo {author} {\bibfnamefont {W.~S.}\ \bibnamefont
  {Lee}}, \bibinfo {author} {\bibfnamefont {E.}~\bibnamefont {Ott}}, \ and\
  \bibinfo {author} {\bibfnamefont {T.~M.}\ \bibnamefont {Antonsen}},\
  }\href@noop {} {\bibfield  {journal} {\bibinfo  {journal} {Phys. Rev. Lett.}\
  }\textbf {\bibinfo {volume} {103}},\ \bibinfo {pages} {044101} (\bibinfo
  {year} {2009})}\BibitemShut {NoStop}%
\bibitem [{\citenamefont {L\"ucken}\ \emph {et~al.}(2013)\citenamefont
  {L\"ucken}, \citenamefont {Pade}, \citenamefont {Knauer},\ and\ \citenamefont
  {Yanchuk}}]{lucken_reduction_2013}%
  \BibitemOpen
  \bibfield  {author} {\bibinfo {author} {\bibfnamefont {L.}~\bibnamefont
  {L\"ucken}}, \bibinfo {author} {\bibfnamefont {J.~P.}\ \bibnamefont {Pade}},
  \bibinfo {author} {\bibfnamefont {K.}~\bibnamefont {Knauer}}, \ and\ \bibinfo
  {author} {\bibfnamefont {S.}~\bibnamefont {Yanchuk}},\ }\href@noop {}
  {\bibfield  {journal} {\bibinfo  {journal} {EPL}\ }\textbf {\bibinfo {volume}
  {103}},\ \bibinfo {pages} {10006} (\bibinfo {year} {2013})}\BibitemShut
  {NoStop}%
\bibitem [{\citenamefont {L\"ucken}\ \emph {et~al.}(2015)\citenamefont
  {L\"ucken}, \citenamefont {Pade},\ and\ \citenamefont
  {Knauer}}]{lucken_classification_2015}%
  \BibitemOpen
  \bibfield  {author} {\bibinfo {author} {\bibfnamefont {L.}~\bibnamefont
  {L\"ucken}}, \bibinfo {author} {\bibfnamefont {J.}~\bibnamefont {Pade}}, \
  and\ \bibinfo {author} {\bibfnamefont {K.}~\bibnamefont {Knauer}},\
  }\href@noop {} {\bibfield  {journal} {\bibinfo  {journal} {SIAM Journal on
  Applied Dynamical Systems}\ }\textbf {\bibinfo {volume} {14}},\ \bibinfo
  {pages} {286} (\bibinfo {year} {2015})}\BibitemShut {NoStop}%
\bibitem [{\citenamefont {Wagemakers}\ and\ \citenamefont
  {Sanju\'an}(2017)}]{wagemakers_2017}%
  \BibitemOpen
  \bibfield  {author} {\bibinfo {author} {\bibfnamefont {A.}~\bibnamefont
  {Wagemakers}}\ and\ \bibinfo {author} {\bibfnamefont {M.}~\bibnamefont
  {Sanju\'an}},\ }\bibfield  {booktitle} {\emph {\bibinfo {booktitle}
  {Scientific Reports}},\ }\href@noop {} {\ \textbf {\bibinfo {volume} {7}},\
  \bibinfo {pages} {2744} (\bibinfo {year} {2017})}\BibitemShut {NoStop}%
\bibitem [{\citenamefont {Goemans}\ and\ \citenamefont
  {Williamson}(1995)}]{Goemans_1995}%
  \BibitemOpen
  \bibfield  {author} {\bibinfo {author} {\bibfnamefont {M.~X.}\ \bibnamefont
  {Goemans}}\ and\ \bibinfo {author} {\bibfnamefont {D.~P.}\ \bibnamefont
  {Williamson}},\ }\href@noop {} {\bibfield  {journal} {\bibinfo  {journal} {J.
  ACM}\ }\textbf {\bibinfo {volume} {42}},\ \bibinfo {pages} {1115} (\bibinfo
  {year} {1995})}\BibitemShut {NoStop}%
\bibitem [{\citenamefont {Newman}(2010)}]{newman2010networks}%
  \BibitemOpen
  \bibfield  {author} {\bibinfo {author} {\bibfnamefont {M.}~\bibnamefont
  {Newman}},\ }\href@noop {} {\emph {\bibinfo {title} {Networks: an
  introduction}}}\ (\bibinfo  {publisher} {Oxford university press},\ \bibinfo
  {year} {2010})\BibitemShut {NoStop}%
\end{thebibliography}
%

\end{document}